\documentclass[11pt]{article}
\usepackage{amsmath,amssymb,amsthm,amsxtra,overpic,bbm,bm,epsfig,subfigure}
\usepackage{color}
\usepackage{comment}
\usepackage{multirow}

\def\thefootnote{\fnsymbol{footnote}}

\begin{document}

\vspace{0.2cm}

\begin{center}
{\Large\bf Majorana Neutrino Masses from Neutrinoless Double-Beta Decays and Lepton-Number-Violating Meson Decays}
\end{center}

\vspace{0.2cm}

\begin{center}
{\bf Jun-Hao Liu $^{a}$} \footnote{E-mail: liujunhao@ihep.ac.cn}
\quad
{\bf Jue Zhang $^{b,~a}$} \footnote{E-mail: zhangjue@ihep.ac.cn}
\quad {\bf Shun Zhou $^{a,~b}$} \footnote{E-mail: zhoush@ihep.ac.cn}
\\
{$^a$Institute of High Energy Physics, Chinese Academy of
Sciences, Beijing 100049, China \\
$^b$Center for High Energy Physics, Peking University, Beijing 100871, China}
\end{center}

\vspace{1.5cm}

\begin{abstract}
The Schechter-Valle theorem states that a positive observation of neutrinoless double-beta ($0\nu \beta \beta$) decays implies a finite Majorana mass term for neutrinos when any unlikely fine-tuning or cancellation is absent. In this note, we reexamine the quantitative impact of the Schechter-Valle theorem, and find that current experimental lower limits on the half-lives of $0\nu \beta \beta$-decaying nuclei have placed a restrictive upper bound on the Majorana neutrino mass $|\delta m^{ee}_\nu| < 7.43 \times 10^{-29}~{\rm eV}$ radiatively generated at the four-loop level. Furthermore, we generalize this quantitative analysis of $0\nu \beta \beta$ decays to that of the lepton-number-violating (LNV) meson decays $M^- \to {M^\prime}^+ + \ell^-_\alpha + \ell^-_\beta$ (for $\alpha$, $\beta$ = $e$ or $\mu$). Given the present upper limits on these rare LNV decays, we have derived the loop-induced Majorana neutrino masses $|\delta m^{ee}_\nu| < 9.7 \times 10^{-18}~{\rm eV}$, $|\delta m^{e\mu}_\nu| <  1.6 \times 10^{-15}~{\rm eV}$ and $|\delta m^{\mu \mu}_\nu| < 1.0 \times 10^{-12}~{\rm eV}$ from $K^- \to \pi^+ + e^- + e^-$, $K^- \to \pi^+ + e^- + \mu^-$ and $K^- \to \pi^+ + \mu^- + \mu^-$, respectively. A partial list of radiative neutrino masses from the LNV decays of $D$, $D_s^{}$ and $B$ mesons is also given.

\end{abstract}

\begin{flushleft}
\hspace{0.8cm} PACS number(s): 11.30.Fs, 12.15.Lk, 14.60.Lm
\end{flushleft}

\def\thefootnote{\arabic{footnote}}
\setcounter{footnote}{0}

\newpage

\section{Introduction}

It remains an open question whether massive neutrinos are Majorana particles, whose antiparticles are themselves~\cite{Majorana:1937vz}. The final answer to this fundamental question will tell us whether the lepton number is conserved or not in nature, and help us explore the origin of neutrino masses.

Currently, the most promising way to answer if massive neutrinos are their own antiparticles is to observe the $0\nu \beta \beta$ decays $N(Z, A) \to N(Z+2, A) + e^- + e^-$, where $Z$ and $A$ stand respectively for the atomic and mass numbers of a nuclear isotope $N(Z, A)$~\cite{GoeppertMayer:1935qp,Furry:1939qr}. Over the last few decades, a great number of dedicated experiments have been carried out to search for this kind of decays~\cite{GomezCadenas:2011it,Doi:1985dx}. So far, we have not observed any positive signals, and a lower bound on the half-life of the implemented nuclear isotope can be drawn from experimental data. The GERDA Phase-I experiment~\cite{Agostini:2013mzu} has disproved the signals of $0\nu\beta \beta$ decays claimed by the Heidelberg-Moscow experiment~\cite{KlapdorKleingrothaus:2004wj}, and the joint lower bound from all the previous $^{76}{\rm Ge}$-based experiments on the half-life turns out to be $T^{1/2}_{0\nu} > 3.0 \times 10^{25}~{\rm yr}$ at the $90\%$ confidence level~\cite{Agostini:2013mzu,KlapdorKleingrothaus:2000sn}. For $^{136}{\rm Xe}$-based experiments, a combined analysis of the EXO-200~\cite{Auger:2012ar} and KamLAND-Zen Phase-I data~\cite{Gando:2012zm} gives rise to a lower bound $T^{1/2}_{0\nu} > 3.4 \times 10^{25}~{\rm yr}$ at the $90\%$ confidence level. More recently, KamLAND-Zen announced their Phase-II result~\cite{KamLAND-Zen:2016pfg}, and improved the lower bound to $T^{1/2}_{0\nu} > 1.07 \times 10^{26}~{\rm yr}$ at the $90\%$ confidence level with both Phase-I and Phase-II data. If neutrino mass ordering is inverted (i.e., $m^{}_3 < m^{}_1 < m^{}_2$), the next-generation $0\nu\beta \beta$ experiments with a few tons of target mass will be able to discover a remarkable signal in the near future~\cite{GomezCadenas:2011it}.

The Schechter-Valle theorem~\cite{Schechter:1981bd} states that a clear signal of $0\nu\beta \beta$ decays will unambiguously indicate a finite Majorana mass of neutrinos, if neither a fine-tuning among parameters nor a cancellation among different contributions is assumed.\footnote{It has been pointed out by Apostolos Pilaftsis that the tree-level parameters can be well chosen to give a vanishing neutrino mass in the type-I seesaw model, while the $0\nu\beta\beta$ decay rate remains nonzero as the nuclear medium effects on quarks may break any intricate cancellation.} Obviously, this theorem signifies the physical importance of searching for $0\nu\beta \beta$ decays experimentally. The quantitative impact of the Schechter-Valle theorem has already been studied by Duerr, Lindner and Merle in Ref.~\cite{Duerr:2011zd}, where it is found that the Majorana neutrino masses implied by the Schechter-Valle theorem are too small to explain neutrino oscillations. Explicitly, assuming one short-range operator to be responsible for $0\nu\beta \beta$ decays, they find that current experimental lower bounds on the half-lives of $0\nu\beta \beta$-decaying isotopes indicate an upper bound on the Majorana neutrino mass $|\delta m^{ee}_\nu| < 5 \times 10^{-28}~{\rm eV}$, where $\delta m^{\alpha \beta}_\nu$ denotes the effective neutrino mass term associated with $\overline{\nu^{}_{\alpha {\rm L}}} \nu^{\rm c}_{\beta {\rm L}}$ for $\alpha, \beta = e, \mu, \tau$. In this paper, we reexamine this problem, and obtain an upper bound $|\delta m^{ee}_\nu| < 7.43\times 10^{-29}~{\rm eV}$ that agrees with the above result from Ref.~\cite{Duerr:2011zd} on the order of magnitude. Furthermore, we generalize the analysis of $0\nu \beta \beta$ decays to that of the LNV rare decays of $B$, $D$ and $K$ mesons. For instance, we obtain $|\delta m^{ee}_\nu| < 9.7  \times 10^{-18}~{\rm eV}$, $|\delta m^{e\mu}_\nu| < 1.6 \times 10^{-15}~{\rm eV}$ and $|\delta m^{\mu \mu}_\nu | < 1.0 \times 10^{-12}~{\rm eV}$ from current upper bounds on the LNV rare decays of $K$ mesons. The radiative Majorana neutrino masses related to other LNV decays are also tabulated. Therefore, we confirm the conclusion from Ref.~\cite{Duerr:2011zd} that although the Schechter-Valle theorem in general implies a tiny Majorana neutrino mass, we have to explore other mechanisms to generate the observed neutrino masses at the sub-eV level.

The remaining part of this work is organized as follows. In Sec.~2, we recall the calculation of Majorana neutrino masses from the four-loop diagram mediated by the effective operator, which is also responsible for the $0\nu \beta \beta$ decays. The generalization to the LNV meson decays is performed in Sec.~3, where the corresponding Majorana masses are computed. Finally, we summarize our main conclusions in Sec.~4.

\section{Majorana Masses from $0\nu \beta \beta$ Decays}

In this section, we present a brief review on the calculation of Majorana neutrino masses radiatively generated from the operator that leads to the $0\nu \beta \beta$ decays, following Ref.~\cite{Duerr:2011zd} closely. Such a calculation can be readily generalized to the case of Majorana neutrino masses induced by the LNV meson decays, as shown in the next section.

\begin{figure}[!t]
\centering
\includegraphics[scale=0.7]{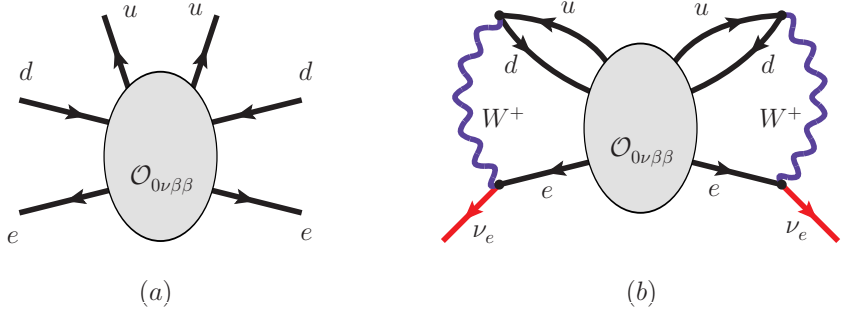}
\caption{The ``ladybird" diagram (a) for the $0\nu\beta \beta$ decays induced by an effective operator ${\cal O}^{}_{0\nu\beta\beta}$, and the ``butterfly" diagram (b) for the corresponding Majorana neutrino mass term $\delta m^{ee}_\nu \overline{\nu^{}_{e{\rm L}}} \nu^{\rm c}_{e{\rm L}}$ generated at the four-loop level~\cite{Schechter:1981bd}.}
\label{fig:0n2b}
\end{figure}

At the elementary-particle level, the $0\nu \beta \beta$ decays can be expressed as $d + d \to u + u + e^- + e^-$, where the up quark $u$, the down quark $d$ and the electron $e^-_{}$ are all massive fermions. If the $0\nu \beta \beta$ decays take place, they can be effectively described by the LNV operator ${\cal O}^{}_{0\nu\beta\beta} = \bar{d} \bar{d} u u e e$, in which the chiralities of charged fermions have been omitted and will be specified later. As already pointed out by Schechter and Valle~\cite{Schechter:1981bd}, this operator will unambiguously result in a Majorana neutrino mass term $\delta m^{ee}_\nu \overline{\nu^{}_{e\rm L}} \nu^{\rm c}_{e\rm L}$. The relevant Feynman diagrams are given in Fig.~\ref{fig:0n2b}. It is worthwhile to notice that quark and charged-lepton masses are indispensable for the Schechter-Valle theorem to be valid, as emphasized in Ref.~\cite{Duerr:2011zd}. In the Standard Model (SM), only left-handed neutrino fields participate in the weak interactions, so the electron masses can be implemented to convert the right-handed electron fields into the left-handed ones, which are then coupled to left-handed neutrino fields via the charged weak gauge boson $W^+$. This does make sense, since the chirality of electrons in the operator ${\cal O}^{}_{0\nu \beta \beta}$ can in general be either left-handed or right-handed. For the same reason, quark masses are also required to realize the hadronic charged-current interactions in the SM. In this case, the operator ${\cal O}^{}_{0\nu\beta\beta}$ in Fig.~\ref{fig:0n2b}(a) can be attached to the left-handed neutrinos through two propagators of $W^+$, leading to the neutrino self-energy diagram in Fig.~\ref{fig:0n2b}(b).

Assuming that $0\nu \beta \beta$ decays are mediated by short-range interactions, one can write down the most general Lorentz-invariant Lagrangian that contains various point-like operators as follows~\cite{Pas:2000vn}
\begin{eqnarray} \label{eq:operators}
\mathcal{L}_{0\nu\beta\beta}^{} &=& \frac{G_{\rm F}^2}{2 m_p^{}} \left (\epsilon_1^{} J J j + \epsilon_2^{} J^{\mu\nu}_{} J_{\mu\nu}^{} j + \epsilon_3^{} J^{\mu}_{} J_{\mu}^{} j + \epsilon_4^{} J^{\mu}_{} J_{\mu\nu}^{} j^{\nu}_{} + \epsilon_5^{} J^{\mu}_{} J j_{\mu}^{} \right) \;,
\end{eqnarray}
where $G_{\rm F}^{} = 1.166 \times 10^{-5}~{\rm GeV}^{-2}$ and $m_p^{} = 938.27~{\rm MeV}$ denote respectively the Fermi constant and the proton mass, and $\epsilon^{}_i$ (for $i = 1, 2, \cdots, 5$) are effective coupling constants. In Eq.~(\ref{eq:operators}), the hadronic currents are defined as~\cite{Pas:2000vn}
\begin{eqnarray}\label{eq:hadronic}
J \equiv \bar{u} (1 \pm \gamma_5^{}) d \; , \quad J^{\mu}_{} \equiv \bar{u} \gamma^{\mu}_{} (1 \pm \gamma_5^{}) d \; , \quad J^{\mu\nu}_{} \equiv \bar{u} \frac{{\rm i}}{2} [\gamma^{\mu}_{}, \gamma^{\nu}_{}] (1 \pm \gamma_5^{}) d \; ,
\end{eqnarray}
while the leptonic currents are given by
\begin{eqnarray}\label{eq:leptonic}
j = \bar{e} (1 \pm \gamma_5^{}) e^{\rm c} \; , \quad j^{\mu}_{} = \bar{e} \gamma^{\mu}_{} (1 \pm \gamma_5^{}) e^{\rm c} \; , \quad j^{\mu\nu}_{} = \bar{e} \frac{{\rm i}}{2} [\gamma^{\mu}_{}, \gamma^{\nu}_{}] (1 \pm \gamma_5^{}) e^{\rm c} \; ,
\end{eqnarray}
where $e^{\rm c}_{} \equiv C\bar{e}^{\rm T}$ with $C \equiv i\gamma^2 \gamma^0$ is the charge-conjugated electron field. According to $C \gamma^{\rm T}_\mu C^{-1} = - \gamma^{}_\mu$ and the fact that fermion fields are Grassmann numbers, one can immediately verify that the tensor leptonic current $j^{}_{\mu \nu}$ automatically vanishes. Different chiralities of hadronic and leptonic currents in Eqs.~(\ref{eq:hadronic}) and (\ref{eq:leptonic}) should be distinguished by the left- and right-handed projection operators $P_{{\rm L},{\rm R}}^{} = (1 \mp \gamma_5^{})/2$. For instance, we have defined $J_{{\rm L},{\rm R}}^{} = 2 \bar{u}  P_{{\rm L}, {\rm R}}^{} d$, and similarly for the other types of currents in Eqs.~(\ref{eq:hadronic}) and (\ref{eq:leptonic}), in which the corresponding subscripts ``L" or ``R" are omitted without causing any confusions. In this connection, the effective coupling constants $\epsilon^{}_i$ should also be regarded as $\epsilon^{\rm xyz}_i$ (for x, y, z = L, R), which are carrying the superscripts for different chiralities of hadronic and leptonic currents.

Given one of the five operators in Eq.~(\ref{eq:operators}), one can set an upper limit on their coupling $\epsilon_i^{}$ by assuming that it is responsible for the $0\nu\beta\beta$ decay and saturates the experimental lower bound on the half-life, as done in Ref.~\cite{Pas:2000vn}. Recently, some of those limits have been recalculated in Ref.~\cite{Duerr:2011zd}, using more recent results for the nuclear matrix elements. The effective coupling constants for the operators $J^{}_{\rm L} J^{}_{\rm L} j^{}_{\rm L}$ and $J^\mu_{\rm R} J^{}_{\mu {\rm R}} j^{}_{\rm L}$ have been found to be $\epsilon^{}_1 < 2\times 10^{-7}$ and $\epsilon^{}_3 < 1.5\times 10^{-8}$ , respectively. Having obtained these couplings, we are then ready to evaluate the induced neutrino mass by inserting the dimension-nine effective operators into the butterfly diagram, as depicted in Fig.~\ref{fig:0n2b}. The authors of Ref.~\cite{Duerr:2011zd} have demonstrated that the operator $J^{}_{\rm L} J^{}_{\rm L} j^{}_{\rm L}$ leads to a vanishing neutrino mass term via the butterfly diagram, while the other one $J_{\rm R}^\mu J_{\mu {\rm R}} j_{\rm L}^{}$ does lead to a tiny Majorana neutrino mass, which will be revisited below.

Now that the operator $\epsilon^{}_3 J_{\rm R}^\mu J_{\mu {\rm R}} j_{\rm L}^{}$ is responsible for the $0\nu \beta \beta$ decays, the radiatively induced Majorana mass term for electron neutrinos can be extracted from the self-energy in Fig.~\ref{fig:0n2b}(b) by setting the external momentum to zero as~\cite{Duerr:2011zd},
\begin{eqnarray} \label{eq:sigmap}
\delta m_{\nu}^{ee} = \frac{128 g^4_{} G_{\rm F}^2 \epsilon_3^{} m_u^2 m_d^2 m_e^2}{m_p^{}} \mathcal{I}_{0\nu\beta\beta}^{} \;,
\end{eqnarray}
where $g$ is the weak gauge coupling, and $m_u^{}, m_d^{}$ and $m_e^{}$ are the up-quark, down-quark and electron masses, respectively. In addition, the loop integral is given by $\mathcal{I}_{0\nu\beta\beta}^{} = \left[\mathcal{I}(m_e^2, m_u^2, m_d^2)\right]^2_{}$ with
\begin{eqnarray}
\mathcal{I}(m_e^2, m_u^2, m_d^2) = \int \frac{{\rm d}^4_{} k_1^{}} {(2\pi)^4} \frac{{\rm d}^4_{} q_1^{}} {(2\pi)^4} \frac{1}{ (k_1^{2} - m_e^2)[(k_1^{} + q_1^{})^2 - m_u^2] (q_1^2 - m_d^2) (k_1^2-M_{W}^2)} \; ,
\end{eqnarray}
where $M_W^{} = 80.4~{\rm GeV}$ is the $W$-boson mass, $k^{}_1$ and $q^{}_1$ stand for the four-momenta of internal particles running in the loop and can be easily identified via the integrand on the right-hand side of Eq.~(5) and from Fig.~\ref{fig:0n2b}(b). To evaluate this integral, we employ the technique for massive two-loop diagrams in Ref.~\cite{Ghinculov:1994sd} and arrive at
\begin{eqnarray}
\mathcal{I}(m_e^2, m_u^2, m_d^2) &=&  \frac{1}{(4\pi^2)^{4+\varepsilon}_{} \mu^{2\varepsilon}_{}} \int_0^1 {\rm d}z ~\mathcal{G} \left((1-z)M_{W}^{2} + z m_e^2, m_d^2, m_u^2; 0 \right) \; ,
\end{eqnarray}
with $\varepsilon \equiv 4-n$ as usually introduced in the dimensional regularization and $\mu$ being the renormalization scale. The relevant function reads as~\cite{Ghinculov:1994sd}
\begin{eqnarray}
\mathcal{G} (m^2_i, m^2_j, m^2_k; k^2) &\equiv& \int ~ \frac{{\rm d}^n p ~ {\rm d}^n q}{(p^2 + m_i^2)^2 [ (q+k)^2 + m_j^2 ] [ (p+q)^2 + m_k^2 ]} \\
&=& \pi^4 (\pi m_i^2)^{n-4} \frac{\Gamma(2-\frac{1}{2}n)}{\Gamma(3-\frac{1}{2}n)} \int_0^1 {\rm d}x \int_0^1 {\rm d}y ~ [x (1-x)]^{\frac{1}{2}n-2} y (1-y)^{2-\frac{1}{2}n} \nonumber \\
& &\times \left\{ \frac{(y^2 \kappa^2 + \eta^2)\Gamma(5-n)}{[y(1-y)\kappa^2 + y + (1-y)\eta^2 ]^{5-n}}  + \frac{n}{2} \frac{\Gamma(4-n)}{[y(1-y)\kappa^2 + y + (1-y)\eta^2 ]^{4-n}} \right \},
\end{eqnarray}
with
\begin{eqnarray}
\eta^2 = \frac{ax+b(1-x)}{x(1-x)}\; , \quad a = \frac{m_j^2}{m_i^2} \; , \quad b = \frac{m_k^2}{m_i^2} \; , \quad \kappa^2 = \frac{k^2}{m_i^2} \;.
\end{eqnarray}
As usual, the integral is expanded with respect to $\varepsilon = 4 - n$ in the limit of $n \to 4$ and the ultraviolet divergences can be separated as inverse powers of $\varepsilon$. Since the loop integral involves the divergent terms proportional to $\varepsilon^{-2}_{}$ and $\varepsilon^{-1}_{}$, we need to keep the terms up to $\varepsilon^2$ in $\mathcal{I}(m^2_e, m^2_u, m^2_d)$, namely,
\begin{eqnarray}
\mathcal{I}(m^2_e, m^2_u, m^2_d) \approx \frac{8.02\times 10^{-5}}{\varepsilon^{2}} - \frac{7.96 \times 10^{-4}}{\varepsilon} + 0.0041 - 0.0146 \varepsilon + 0.040 \varepsilon^2 + \mathcal{O}(\varepsilon^3) \; ,
\end{eqnarray}
so as to obtain all the finite parts of ${\cal I}^{}_{0\nu \beta \beta}$.
In our numerical calculations, we have adopted the renormalization scale of $\mu = 100~{\rm MeV}$, which is a characteristic scale of typical energy transfer in nuclear processes. The other implemented parameters can be found in Table~\ref{tb:constants}. In the scheme of minimal subtraction, we finally get the induced neutrino mass from Eq.~(\ref{eq:sigmap}) as
\begin{eqnarray}
|\delta m_{\nu}^{ee}| < 7.43 \times 10^{-29} ~{\rm eV} \; ,
\end{eqnarray}
which agrees with the result $|\delta m^{ee}_\nu| < 5 \times 10^{-28}~{\rm eV}$ from Ref.~\cite{Duerr:2011zd} on the order of magnitude.\footnote{Here we perform a careful treatment on the renormalization scale in the evaluation of the loop integral, and that leads to the above minor discrepancy between two numerical results. At this point, we are grateful to Dr. Michael Duerr for kind communications regarding the evaluation of the loop integral.}  Since this mass is extremely small, one has to implement other mechanisms to account for neutrino masses. In this sense, the main conclusion in Ref.~\cite{Duerr:2011zd} is still valid that the Schechter-Valle theorem is qualitatively correct, but quantitatively irrelevant for the neutrino mass-squared differences required for neutrino oscillation experiments.
\begin{table}
\centering
\begin{tabular}{ l l l l}
\hline \hline
 Quark masses & $m^{}_u = 2.3 ~{\rm MeV}$ & $m^{}_d = 4.8 ~{\rm MeV}$ & $m^{}_s = 95 ~{\rm MeV}$\\
 ~ & $m^{}_c = 1.27 ~{\rm GeV}$ & $m^{}_b = 4.2 ~{\rm GeV}$ & ~ \\
\hline
 Lepton masses & $m^{}_e = 0.511 ~{\rm MeV}$ & $m^{}_\mu = 105.7 ~{\rm MeV}$ & ~ \\
\hline
 Meson masses & $m^{}_\pi = 139.6 ~{\rm MeV}$ & $m^{}_\rho = 775.3 ~{\rm MeV}$ & $m^{}_K = 493.7 ~{\rm MeV}$ \\
 ~  & $m^{}_{K^*} = 891.7 ~{\rm MeV}$ & $m^{}_D = 1869.6 ~{\rm MeV}$ & $m^{}_{D_s} = 1968.3 ~{\rm MeV}$  \\
 ~ &  $m^{}_B = 5279.3 ~{\rm MeV}$ & ~ & ~ \\
\hline
 Decay constants & $f^{}_\pi = 135 ~{\rm MeV}$ & $f^{}_\rho = 209 ~{\rm MeV}$ & $f^{}_K = 160 ~{\rm MeV}$ \\
 ~ & $f^{}_{K^*} = 218 ~{\rm MeV}$ & $f^{}_D = 222.6 ~{\rm MeV}$ & $f^{}_{D_s} = 280.1 ~{\rm MeV}$ \\
 ~ & $f^{}_B = 216 ~{\rm MeV}$ & ~ & ~  \\
 \hline
\multirow{2}{1.4in}{CKM mixing angles and Dirac phase} & $\theta_{12}^{} = 0.227$ & $\theta_{23} = 0.041$ & $\theta_{13} = 3.6 \times 10^{-3}$ \\
& $\delta = 1.25$ & &  \\
\hline \hline
\end{tabular}
\vspace{0.2cm}
\caption{Particle masses, decay constants of mesons, CKM mixing angles and Dirac phase that are used in the evaluation of radiatively-generated neutrino masses~\cite{Agashe:2014kda}.}
\label{tb:constants}
\end{table}

\section{Majorana Masses from LNV Meson Decays}

Then, we consider the LNV meson decays $M_i^- \rightarrow M_f^+ \ell_\alpha^- \ell_\beta^-$, where $M_i^-$ and $M_f^+$ are the initial and final charged mesons, while the emitted same-sign charged leptons with flavors $\alpha$ and $\beta$ are denoted by $\ell_{\alpha}^-$ and $\ell_{\beta}^-$, respectively. These processes have been extensively discussed in the presence of heavy Majorana neutrinos or a Higgs triplet~\cite{meson_decays}. If the LNV meson decays are observed experimentally, we assume that these processes are caused by some short-range interactions and can be described by a number of Lorentz-invariant operators of dimension-nine. However, to carry out an order-of-magnitude estimate of the induced Majorana neutrino masses, one can simply consider just one operator so long as it contributes dominantly. The main idea is to generalize the analysis for $0\nu\beta \beta$ decays to the LNV meson decays. For instance, we take the operator
\begin{eqnarray}
\mathcal{L}_{\rm MD}^{} = \varepsilon_{\alpha\beta}^{} \frac{G_{\rm F}^2}{2 m_{p}^{}} J_{\rm R}^\mu J_{\mu {\rm R}}^\prime j_{\rm L}^{} \; ,
\end{eqnarray}
where $\varepsilon_{\alpha\beta}^{}$ (for $\alpha, \beta$ = $e, \mu$) are real dimensionless couplings, and the hadronic and leptonic currents are defined similarly as before, i.e., $J_{\mu {\rm R}}^{(\prime)} = 2\overline{U^{(\prime)}} \gamma_\mu^{} P_{\rm R}^{} D^{(\prime)}$ and $j_{\rm L}^{} = 2\overline{\ell_\alpha^{}} P_{\rm L}^{} \ell_\beta^{\rm c}$. Here $U^{(\prime)}$ and $D^{(\prime)}$ are generic up- and down-type quark fields, and we have distinguished two possibly different hadronic currents by a prime symbol. For instance, in the decay of $B^-_{} \rightarrow D^+_{} e^-_{} \mu^-_{}$, the two hadronic currents are $\bar{c}\gamma_\mu^{}P_{\rm R}^{} d$ and $\bar{u}\gamma_\mu^{} P_{\rm R}^{} b$, with $b$ being the bottom-quark field, while the leptonic current is $\overline{e} P_{\rm L}^{} \mu^{\rm c}_{}$. It should be noticed that the results will also be valid for the CP-conjugated channel $M_i^+ \rightarrow M_f^- \ell_\alpha^+ \ell_\beta^+$ if the CP violation in these LNV decays is negligible.

Given the operator in Eq.~(12), it is straightforward to write down the Feynman amplitude for the LNV meson decay $M_i^-(p^{}_i) \rightarrow M_f^+(p^{}_f) + \ell_\alpha^-(p^{}_\alpha) + \ell_\beta^-(p^{}_\beta)$ as
\begin{eqnarray} \label{eq:matrix_element}
{\rm i} \mathcal{M} = {\rm i} \varepsilon_{\alpha\beta}^{} \frac{G_{\rm F}^2}{2 m_{p}^{}}~ 8\langle M_f^+ (p_f^{}) \ell_\alpha^-(p_\alpha^{}) \ell_\beta^- (p_\beta^{}) |~ \overline{U} \gamma^\mu_{} P_{\rm R}^{} D \overline{U^\prime} \gamma_\mu^{} P_{\rm R}^{} D^\prime_{}  \overline{\ell_\alpha^{}} P_{\rm L}^{} \ell_\beta^c ~| M_i^- (p_i^{}) \rangle \; ,
\end{eqnarray}
where the four-momenta of initial and final states have been specified explicitly. Since the initial and final mesons are bound states, the hadronic processes involving them in general cannot be calculated perturbatively. However, in our case, we assume that the hadronic interactions can be factorized out and related to the leptonic meson decay constants, which are defined as follows~\cite{Agashe:2014kda}
\begin{eqnarray}
\langle 0 | \bar{q}\gamma_\mu^{} \gamma_5^{} q^\prime  | P(p) \rangle = -ip_\mu^{} f_P^{} \; , \quad \langle 0 | \bar{q} \gamma_\mu^{} q^\prime |V(p) \rangle = \epsilon_\mu^{} m_V^{} f_V^{} \; ,
\end{eqnarray}
where $P$ and $V$ respectively denote pseudoscalar and vector mesons, $\epsilon_\mu^{}$ the polarization vector for $V$, and $m_V^{}$ the vector-meson mass. For the relevant decay constants $f_{P}^{}$ and $f^{}_V$, we adopt their numerical values from Ref.~\cite{Agashe:2014kda} and list them in Table~\ref{tb:constants}.

For illustration, we first deal with the decay rates of pseudoscalar mesons. The results for the vector-meson decays can be similarly obtained, and will be given later in this section. With the help of the decay constants, the square of amplitude in Eq.~(\ref{eq:matrix_element}) can be reduced to
\begin{eqnarray}
|\mathcal{M}|^2 = \frac{G_{\rm F}^4}{4 m_p^4} \varepsilon_{\alpha\beta}^2 f_i^2 f_f^2 (p_i^{} \cdot p_f^{})^2 (p_\alpha^{} \cdot p_\beta^{}) \; ,
\end{eqnarray}
which will be inserted into the standard formula of the differential rate for three-body decays and lead to
\begin{eqnarray}
\frac{d\Gamma}{ds} = \frac{1}{2 m_i^{}} \int \frac{d^3 p^{}_f}{(2\pi)^3}\frac{1}{2 E_f^{}} \int \frac{d^3 p_\alpha^{}}{(2\pi)^3} \frac{1}{2E_\alpha^{}} \int \frac{d^3 p_{\beta}^{}}{(2\pi)^3} \frac{1}{2E_{\beta}^{}} |\mathcal{M}|^2 (2\pi)^4 \delta^4(q-p_\alpha^{} - p_\beta^{}) \delta[q^2 - (p_i^{} - p_f^{})^2] \; ,
\end{eqnarray}
where $E_{f,\alpha,\beta}^{}$ are the energies of final-state particles, and $s \equiv q^2$ is the invariant momentum square transferred to leptons so that the condition $(m_\alpha^{} + m_\beta^{})^2 \leq s \leq (m_i^{} - m_f^{})^2$ is satisfied. Here $m_{i,f,\alpha,\beta}^{}$ stand for the masses of the initial- and final-state particles. After a direct evaluation of the integral, we obtain
\begin{eqnarray}
\frac{d\Gamma}{ds} = \frac{C_{\alpha\beta}^{}}{2}\frac{\varepsilon_{\alpha\beta}^2 G_{\rm F}^4 f_i^2 f_f^2}{16 (4\pi)^3 m_i^3 m_{p}^2} \lambda^{1/2}_{} (s, m_\alpha^2, m_\beta^2) \lambda^{1/2}_{} (s, m_i^2, m_f^2) \frac{(s-m_i^2-m_f^2)^2 (s-m_\alpha^2 - m_\beta^2)}{s} \; ,
\end{eqnarray}
where $C_{\alpha\beta}^{} = 1 (2)$ for $\alpha = \beta$ ($\alpha \neq \beta$), and $\lambda(a,b,c) \equiv (a-b-c)^2 - 4bc$ is the K\"allen function. Then, the LNV decay rates of vector mesons can be derived in a similar way, and the final results turn out to be
\begin{eqnarray}
\frac{d\Gamma}{ds} = \frac{C_{\alpha\beta}^{}}{2}\frac{\varepsilon_{\alpha\beta}^2 G_{\rm F}^4 f_i^2 f_f^2}{16 (4\pi)^3 m_i^3 m_{p}^2} \lambda^{1/2}_{} (s, m_\alpha^2, m_\beta^2) \lambda^{3/2}_{} (s, m_i^2, m_f^2) \frac{ (s-m_\alpha^2 - m_\beta^2)}{s} \; .
\end{eqnarray}
Finally, the partial decay width $\Gamma$ can be computed by integrating the differential one $d\Gamma/ds$ over the allowed range of $s$. By comparing between current experimental bounds on the LNV rare decays from Ref.~\cite{Agashe:2014kda} and theoretical predictions, one can extract the upper limits on the corresponding coupling constants $\varepsilon_{\alpha\beta}^{}$. In Table \ref{tb:numass_meson}, we list such upper limits for a number of LNV meson-decay processes, and those numerical values will be used to compute the neutrino masses radiatively generated at the four-loop level, as shown in Fig.~\ref{fig:0n2b}(b).

Since we have chosen the operator in Eq.~(12) for the LNV meson decays, which resembles well the one $\epsilon^{}_3 J^\mu_{\rm R} J^{}_{\mu {\rm R}} j^{}_{\rm L}$ for $0\nu\beta \beta$ decays in the previous section,  the calculation of generated Majorana neutrino mass terms from LNV meson decays follows closely that in the case of $0\nu\beta\beta$ decays. The only difference is the presence of two possibly different lepton flavors and different hadronic currents, which bring the CKM matrix elements into the calculation. In the case where $U \neq U^\prime_{}$ and $D \neq D^\prime_{}$ do not hold simultaneously, a straightforward evaluation of a similar butterfly diagram leads to an induced neutrino mass $\delta m^{\alpha\beta}_{\nu}$ for $\alpha$ and $\beta$ lepton flavors, namely,
\begin{eqnarray}
\delta m^{\alpha\beta}_{\nu} &=& \frac{64 g^4_{} {\rm G}_{\rm F}^2  \varepsilon_{\alpha\beta}^{} m_U^{} m_D^{} m_{U^\prime}^{} m_{D^\prime}^{} m_{\alpha}^{} m_{\beta}^{}}{C^{}_{U U^\prime} C^{}_{D D^\prime} m_{\rm p}^{}}
\left[  V_{U D}^{*}  V_{U^\prime D^\prime}^{*} \cdot \mathcal{I}(m_\alpha^2, m_U^2, m_D^2) \cdot \mathcal{I}(m_\beta^2, m^2_{U^\prime}, m^2_{D^\prime}) + (\alpha \leftrightarrow \beta)\right] \; , \nonumber \\
\end{eqnarray}
where $V_{U^{(\prime)} D^{(\prime)}}$ is the CKM matrix element, $C^{}_{U U^\prime}$ and $C^{}_{D D^\prime}$ follow the same definition of $C^{}_{\alpha \beta}$ below Eq.~(17), and the loop integral $\mathcal{I}$ is the same as that introduced in Eq.~(6). On the other hand, when $U \neq U^\prime_{}$ and $D \neq D^\prime_{}$ are both present, we obtain
\begin{eqnarray}
\delta m^{\alpha\beta}_{\nu} &=& 16 g^4_{} {\rm G}_{\rm F}^2  \varepsilon_{\alpha\beta}^{} m_U^{} m_D^{} m_{U^\prime}^{} m_{D^\prime}^{} m_{\alpha}^{} m_{\beta}^{} m_{\rm p}^{-1}
\left[  V_{U D}^{*}  V_{U^\prime D^\prime}^{*} \cdot \mathcal{I}(m_\alpha^2, m_U^2, m_D^2) \cdot \mathcal{I}(m_\beta^2, m^2_{U^\prime}, m^2_{D^\prime}) \right. \nonumber \\
& & \qquad \qquad \left . +   V_{U^\prime D}^{*}  V_{U D^\prime}^{*} \cdot \mathcal{I}(m_\alpha^2, m^2_{U^\prime}, m_D^2) \cdot \mathcal{I}(m_\beta^2, m^2_{U}, m^2_{D^\prime}) + (\alpha \leftrightarrow \beta)\right] \; .
\end{eqnarray}
Using numerical values of quark and lepton masses, CKM mixing angles $\theta_{ij}^{}$ (for $ij=12, 13, 23$) and Dirac phase $\delta$ in Table \ref{tb:constants}, we have tabulated the Majorana neutrino masses implied by various types of LNV meson decays in Table~\ref{tb:numass_meson}.

As one can observe from Table~\ref{tb:numass_meson}, depending on the current experimental limits, the values of Majorana neutrino masses from LNV meson decays can be quite different, spanning over many orders of magnitude. The LNV meson decays may indicate Majorana neutrino mass terms $\delta m^{e\mu}_\nu$ and $\delta m^{\mu \mu}_\nu$, which cannot be obtained from $0\nu\beta\beta$ decays. For instance, if the LNV decays $K^- \to \pi^+ e^- \mu^-$ and $K^- \to \pi^+ \mu^- \mu^-$ are observed,  we arrive at $|\delta m^{e\mu}_\nu| \sim  1.6 \times 10^{-15}~{\rm eV}$ and $|\delta m^{\mu \mu}_\nu| \sim 1.0 \times 10^{-12}~{\rm eV}$, which are still far below the required masses from neutrino oscillation experiments.

\section{Summary}

Whether massive neutrinos are Majorana or Dirac particles remains an unsolved fundamental problem in particle physics. According to the Schechter-Valle theorem, if the $0\nu\beta\beta$ decays $N(Z, A) \to N(Z+2, A) + e^- + e^-$ are observed in future experiments, one can claim that neutrinos do have Majorana masses. In this short note, we have revisited the quantitative impact of the Schechter-Valle theorem and shown that the Majorana neutrino mass radiatively generated at the four-loop level is $|\delta m^{ee}_\nu| < 7.43\times 10^{-29}~{\rm eV}$. Furthermore, a similar analysis has been performed for the LNV meson decays $M^- \to M^{\prime +} + \ell^-_\alpha + \ell^-_\beta$, from which the upper bounds $|\delta m^{ee}_\nu| < 9.7 \times 10^{-18}~{\rm eV}$, $|\delta m^{e\mu}_\nu| < 1.6 \times 10^{-15}~{\rm eV}$ and $|\delta m^{\mu \mu}_\nu| < 1.0 \times 10^{-12}~{\rm eV}$ can be derived. A list of radiative neutrino masses from other LNV rare decays of $D$ and $B$ mesons is also given.

Therefore, even if the $0\nu\beta \beta$ decays or the LNV meson decays are detected and the decay rates are close to current upper bounds, we have to invoke some other mechanisms to produce sub-eV neutrino masses, which can be of either Dirac or Majorana nature. In the former case, massive neutrinos should be pseudo-Dirac particles, since a small Majorana mass is implied by the LNV decays. In the latter case, compared to the sub-eV neutrino masses at the leading order, the radiative Majorana masses can be neglected.

\section*{Acknowledgments}
One of the authors (S.Z.) is grateful to Alexander Merle and Apostolos Pilaftsis for helpful discussions, and to the Mainz Institute for Theoretical Physics (MITP) for its hospitality and its partial support during the completion of this work, which has also been supported in part by the National Recruitment Program for Young Professionals and by the CAS Center for Excellence in Particle Physics (CCEPP).

\begin{table}
\centering
\begin{tabular}{l  c  c  c }
\hline \hline
Decay modes & Branching ratios & Upper bounds on $\varepsilon_{\alpha\beta}^{}$ & Upper bounds on $|\delta m^{\alpha\beta}_{\nu}|$ (eV) \\
\hline
$K^- \rightarrow \pi^+ e^- e^-$ & $ < 6.4 \times 10^{-10}$ & $9.0 \times 10^2$ & $9.7 \times 10^{-18}$\\
$K^- \rightarrow \pi^+ \mu^- \mu^-$ & $ < 1.1 \times 10^{-9}$ & $2.2 \times 10^3$ & $1.0 \times 10^{-12}$\\
$K^- \rightarrow \pi^+ e^- \mu^-$ & $ < 5.0 \times 10^{-10}$ & $7.3 \times 10^2$ & $1.6 \times 10^{-15}$\\
\hline
$D^- \rightarrow \pi^+ e^- e^-$ & $ < 1.1 \times 10^{-6}$ & $2.4 \times 10^4$ & $7.3 \times 10^{-15}$ \\
$D^- \rightarrow \pi^+ \mu^- \mu^-$ & $ < 2.2 \times 10^{-8}$ & $3.5 \times 10^3$ & $4.6 \times 10^{-11}$ \\
$D^- \rightarrow \pi^+ e^- \mu^-$ & $ < 2.0 \times 10^{-6}$ & $2.4 \times 10^4$ & $1.5 \times 10^{-12}$ \\
$D^- \rightarrow \rho^+ \mu^- \mu^-$ & $ < 5.6 \times 10^{-4}$ & $1.0 \times 10^6$ & $1.3 \times 10^{-8}$\\
$D^- \rightarrow K^+ e^- e^-$ & $ < 9 \times 10^{-7}$ & $2.1 \times 10^4$ & $2.5 \times 10^{-13}$ \\
$D^- \rightarrow K^+ \mu^- \mu^-$ & $ < 1.0 \times 10^{-5}$ & $7.2 \times 10^4$ & $3.7 \times 10^{-8}$\\
$D^- \rightarrow K^+ e^- \mu^-$ & $ < 1.9 \times 10^{-6}$ & $2.2 \times 10^4$ & $5.5 \times 10^{-11}$\\
$D^- \rightarrow K^{* +} \mu^- \mu^-$ & $ < 8.5 \times 10^{-4}$ & $1.7 \times 10^6$ & $8.7 \times 10^{-7}$\\
\hline
$D_s^- \rightarrow \pi^+ e^- e^-$ & $< 4.1 \times 10^{-6}$ & $4.5 \times 10^4$ & $5.5 \times 10^{-13}$ \\
$D_s^- \rightarrow \pi^+ \mu^- \mu^-$ & $< 1.2 \times 10^{-7}$ & $7.9 \times 10^3$ & $4.1 \times 10^{-9}$\\
$D_s^- \rightarrow \pi^+ e^- \mu^-$ & $< 8.4 \times 10^{-6}$ & $4.6 \times 10^4$ & $1.2 \times 10^{-10}$\\
$D_s^- \rightarrow K^+ e^- e^-$ & $< 5.2 \times 10^{-6}$ & $4.7 \times 10^4$ & $5.6 \times 10^{-12}$\\
$D_s^- \rightarrow K^+ \mu^- \mu^-$ & $< 1.3 \times 10^{-5}$ & $7.7 \times 10^4$ & $3.9 \times 10^{-7}$\\
$D_s^- \rightarrow K^+ e^- \mu^-$ & $< 6.1 \times 10^{-6}$ & $3.7 \times 10^4$ & $8.9 \times 10^{-10}$ \\
$D_s^- \rightarrow K^{* +} \mu^- \mu^-$ & $< 1.4 \times 10^{-3}$ & $1.8 \times 10^6$ & $9.1 \times 10^{-6}$ \\
\hline
$B^- \rightarrow \pi^+ e^- e^-$ &  $ < 2.3 \times 10^{-8}$ & $7.6 \times 10^1$  &  $5.7 \times 10^{-19}$ \\
$B^- \rightarrow \pi^+ \mu^- \mu^-$ & $< 1.3 \times 10^{-8}$ & $5.7\times 10^1$ & $1.8 \times 10^{-14}$ \\
$B^- \rightarrow \pi^+ e^- \mu^-$ & $ < 1.5 \times 10^{-7}$ & $1.4\times 10^2$ & $2.1 \times 10^{-16}$ \\
$B^- \rightarrow \rho^+ e^- e^-$ & $ < 1.7 \times 10^{-7}$ & $1.5\times 10^2$ & $1.2 \times 10^{-18}$\\
$B^- \rightarrow \rho^+ \mu^- \mu^-$ & $ < 4.2 \times 10^{-7}$ & $2.4\times 10^2$ & $7.5 \times 10^{-14}$ \\
$B^- \rightarrow \rho^+ e^- \mu^-$ & $ < 4.7 \times 10^{-7}$ & $1.8\times 10^2$ & $2.8 \times 10^{-16}$\\
$B^- \rightarrow K^+ e^- e^-$ & $ < 3.0 \times 10^{-8}$ & $7.4\times 10^1$ & $2.5 \times 10^{-18}$\\
$B^- \rightarrow K^+ \mu^- \mu^-$ & $ < 4.1 \times 10^{-8}$ & $8.7\times 10^1$ & $1.3 \times 10^{-13}$\\
$B^- \rightarrow K^+ e^- \mu^-$ & $ < 1.6 \times 10^{-7}$ &  $1.2\times 10^2$ & $8.6 \times 10^{-16}$\\
$B^- \rightarrow K^{* +} e^- e^-$ & $ < 4.0 \times 10^{-7}$ & $2.4\times 10^2$ & $8.1 \times 10^{-18}$ \\
$B^- \rightarrow K^{* +} \mu^- \mu^-$ & $ < 5.9 \times 10^{-7}$ & $2.9\times 10^2$ & $4.2 \times 10^{-13}$\\
$B^- \rightarrow K^{* +} e^- \mu^-$ & $ < 3.0 \times 10^{-7}$ & $1.5 \times 10^2$ & $1.0 \times 10^{-15}$ \\
$B^- \rightarrow D^+ e^- e^-$ & $ < 2.6 \times 10^{-6}$ & $6.3\times 10^2$ & $1.5 \times 10^{-14}$\\
$B^- \rightarrow D^+ \mu^- \mu^-$ & $ < 6.9 \times 10^{-7}$ & $3.3\times 10^2$ & $3.3 \times 10^{-10}$\\
$B^- \rightarrow D^+ e^- \mu^-$ & $ < 1.8 \times 10^{-6}$ & $3.7\times 10^2$ & $1.8 \times 10^{-12}$ \\
$B^- \rightarrow D^{ +}_s \mu^- \mu^-$ & $ < 5.8 \times 10^{-7}$ & $2.5\times 10^2$ & $1.3 \times 10^{-9}$ \\
\hline \hline
\end{tabular}
\vspace{0.4cm}
\caption{A partial list of the LNV decays of $K$, $D$, $D^{}_s$ and $B$ mesons and current experimental constraints on the branching ratios~\cite{Agashe:2014kda}. The upper bounds on the coefficients $\varepsilon^{}_{\alpha \beta}$ and those on radiative neutrino masses $|\delta m^{\alpha \beta}_\nu|$ (for $\alpha, \beta$ = $e, \mu$) are given in the last two columns.}
\label{tb:numass_meson}
\end{table}

\end{document}